\documentclass{ws-p8-50x6-00}
\usepackage[dvips]{color}
\usepackage{epsfig}

\def\bmath#1{\mbox{\boldmath $#1$}}
\def\scbmath#1{\mbox{\boldmath $\scriptstyle #1$}}
\def\ds{\displaystyle}

\begin{document}

\title{\bf Dynamical nature of the nuclear pseudospin and its isospin asymmetry}


\author{R. Lisboa, M. Malheiro, A. Delfino}
\address{Instituto de F\'\i sica, Universidade Federal Fluminense, 
Niter\'oi, Brazil}
\author{P. Alberto, M. Fiolhais}
\address{Departamento de F\'\i sica and Centro de F\'\i sica Computacional, 
Universidade de Coimbra, P-3004-516 Coimbra, Portugal}
\author{M. Chiapparini}
\address{Departamento de F\'\i sica Te\'orica, Universidade do Estado do Rio
de Janeiro, Brazil}
\maketitle
\noindent

\abstracts{
Pseudospin symmetry in nuclei is investigated by solving the
Dirac equation with Woods-Saxon scalar and vector radial potentials.
We relate the pseudospin interaction with a pseudospin-orbit term
in a Schroedinger-like equation for the lower component
of the Dirac spinor. We show that this term gives a large contribution to the 
energy splittings of pseudospin partners, so that the near pseudospin degeneracy
arises from a significant cancellation among the different terms in that equation. 
This is a manifestation of the dynamical character of this symmetry in the nucleus.
We analyze the isospin dependence of the pseudospin symmetry in a nuclear isotope chain
by including a vector-isovector potential $V_\rho$ and a Coulomb potential and conclude that
$V_\rho$ gives the main contribution to the observed pseudospin isospin asymmetry.}

\begin{section}{Introduction}

\noindent Pseudospin symmetry was introduced in the late 60's to account for a 
degeneracy
in single-particle nucleon levels of certain nuclei, with
quantum numbers ($n$, $\ell$, $j= \ell + 1/2$) and ($n-1$, $\ell+2$, $j= \ell + 
3/2$) \cite{kth,aa}. These levels have the same quantum number $\tilde{\ell} = 
\ell +1$, called ``pseudo'' orbital angular momentum, and a ``pseudo'' spin quantum number,
$\tilde s = 1/2$. In general, for a $n\ell j$ state, we have $\tilde\ell=\ell\pm 
1$ if $j=\ell\pm 1/2$. Pseudospin symmetry is exact when
doublets with $j = \tilde{\ell}\ \pm \tilde s$ are degenerate.
Since then the origin of pseudospin symmetry in nuclei has been a
subject of intense theoretical research.

Recently, Ginocchio \cite {gino} uncovered the relativistic character of the symmetry,
noting that the pseudo-orbital angular momentum is the orbital
angular momentum of the lower component of the Dirac spinor. This is an exact 
SU(2) symmetry for the Dirac Hamiltonian \cite{levi} with an
attractive scalar potential, $S$, and a repulsive vector potential $V$,
when they have the same magnitude, i.e.,  $\Sigma=S+V=0$.
This would explain why there are quasi-degenerate pseudospin doublets in 
nuclei, since in relativistic mean field (RMF) models nuclear saturation arises from
an extensive cancellation between a large attractive scalar potential and a large
repulsive vector potential. However, since $\Sigma$ acts as a binding 
potential, one cannot set $\Sigma=0$ in the nucleus, because then there would not 
exist any bound states. Other authors have studied the role of pseudospin-orbit
coupling in pseudospin symmetry \cite{arima,marcos2}. 


\end{section}

\begin{section}{Dynamical pseudospin symmetry and isospin asymmetry}

\noindent
Pseudospin symmetry is related to an invariance under a SU(2)
transformation of the Dirac Hamiltonian with 
central scalar $S$ and vector $V$ potentials
\begin{eqnarray}
H = \mbox{\boldmath $\alpha\cdot p$}
+ \beta (m  + S) + V  \ ,
\label {dirac}
\end{eqnarray}
whose generators are \cite{levi}
$S_i=s_i\,{1\over 2}(1-\beta)+\tilde s_i
\,{1\over 2}(1+\beta)$ where
$\tilde s_i={\scbmath\sigma\cdot\scbmath p\over p} s_i
{\scbmath\sigma\cdot\scbmath p\over p}$, $s_i={\sigma_i/2}$.
The commutator of $S_i$ with (\ref{dirac}) is zero if
${\Sigma = 0}$ or $ {d\Sigma\over dr}=0$ \cite{pmmdm_prl_prc}.

The Dirac equation $H\Psi=\epsilon\Psi$ can be written as a pair of 
second-order differential 
equations for the upper and lower components. Setting
$\Delta=V-S$, $V$ and $S$ being radial functions, and defining $\epsilon=E+m$,
  we have
\begin{eqnarray}
p^2\Psi_{+}&=&\frac{\Delta'}{E+2m-\Delta}\bigg(\frac{\partial\hfill}{\partial 
r}-
\frac{1}{r} \bmath{\sigma}\cdot\bmath{L}\bigg)\Psi_{+}
+(E+2m-\Delta)(E-\Sigma)\Psi_{+}
\\
p^2\Psi_{-}&=&\frac{\Sigma'}{E-\Sigma}\bigg(\frac{\partial\hfill}{\partial r}-
\frac{1}{r} \bmath{\sigma}\cdot\bmath{L}\bigg)\Psi_{-}
+(E+2m-\Delta)(E-\Sigma)\Psi_{-}\ ,
\label{psiminus}
\end{eqnarray}
where $\Psi_{\pm}=[(1\pm\beta)/2]\Psi$ are the upper and lower 
components and the primes denote derivatives with
respect to $r$. The $\bmath\sigma\cdot\bmath{L}$ term in (\ref{psiminus}) is the 
pseudospin-orbit term. 

We solve the Dirac equation with the Hamiltonian (\ref{dirac}) with
scalar and vector potentials with Woods-Saxon shape:
$
U(r)=U_0/ \big(1+\exp [(r-R)/a]\big)
$.
We fitted the parameters of this potential to the neutron spectra of $^{208}$Pb, 
obtaining the values 
$R=7$ fm, $\Delta_0=650$ MeV, $\Sigma_0=-66$ MeV and $a=0.6$ fm. We then
varied the diffuseness, radius and $\Sigma_0$ separetely, keeping all the
other parameters fixed and showed that the splittings 
of the pseudospin doublets vary notably,
namely they decrease as $|\Sigma_0|$ decreases and diffuseness increases, 
sometimes reversing sign.
The details can be found in ref.~\cite{pmmdm_prl_prc}.

The effect of the pseudospin-orbit term on pseudospin energy splittings
can be assessed by writing (\ref{psiminus}) as a 
Schroedinger-like equation, dividing it by an energy- and $r$-dependent effective mass 
$m^*=(E+2m-\Delta)/2$ and taking the expectation value of each term relative to $\Psi_-$:  
\begin{equation}
\bigg\langle \frac{p^2}{2m^*}\bigg\rangle+\big\langle V_{\rm PSO}\big\rangle +
\big\langle V_{\rm D}\big\rangle+\langle\Sigma\rangle=E \ ,
\label{aveg_schroed}
\end{equation}
where
\begin{eqnarray}
\label{kinetic}
\bigg\langle \frac{p^2}{2m^*}\bigg\rangle&=&
\ds\int\Psi_-^\dagger\frac{p^2}{2 m^*}\Psi_-\,{\rm d}^3{\bmath r}\bigg/
\int\Psi_-^\dagger\Psi_-\,{\rm d}^3{\bmath r}\ ,\\
\label{pso}
\big\langle V_{\rm PSO}\big\rangle&=&
{\rm P\ }\int\Psi_-^\dagger\frac{1}{2 m^*}\frac{\Sigma'}{E-\Sigma}
\frac{1}{r} \bmath{\sigma}\cdot\bmath{L}\Psi_-{\rm d}^3{\bmath r}
\bigg/
\int\Psi_-^\dagger\Psi_-\,{\rm d}^3{\bmath r}\\
\label{darwin}
\big\langle V_{\rm D}\big\rangle&=&
-{\rm P}\ds\int\Psi_-^\dagger\frac{1}{2 m^*}\frac{\Sigma'}{E-\Sigma}
\frac{\partial\Psi_-}{\partial r}\,{\rm d}^3{\bmath r}\bigg/
\int\Psi_-^\dagger\Psi_-\,{\rm d}^3{\bmath r}\ ,\\
\label{sigma_aver}
\langle\Sigma\rangle&=&
\ds\int\Psi_-^\dagger\Sigma\Psi_-\,{\rm d}^3{\bmath r}\bigg/
\int\Psi_-^\dagger\Psi_-\,{\rm d}^3{\bmath r}\ ,
\end{eqnarray}
where {\textquoteleft P\textquoteright} denotes the principal value of the 
integral. These terms can be identified, respectively, as a kinetic term, a 
pseudospin-orbit term,
 a potential term related to what is sometimes called Darwin term
and the mean value of the $\Sigma$ potential with respect to the lower 
component, $\Psi_-$.
\vskip-.2cm
\begin{figure}[hbt]
\parbox[t]{5.5cm}{
\begin{center}
\includegraphics[clip=on,width=5.5cm]{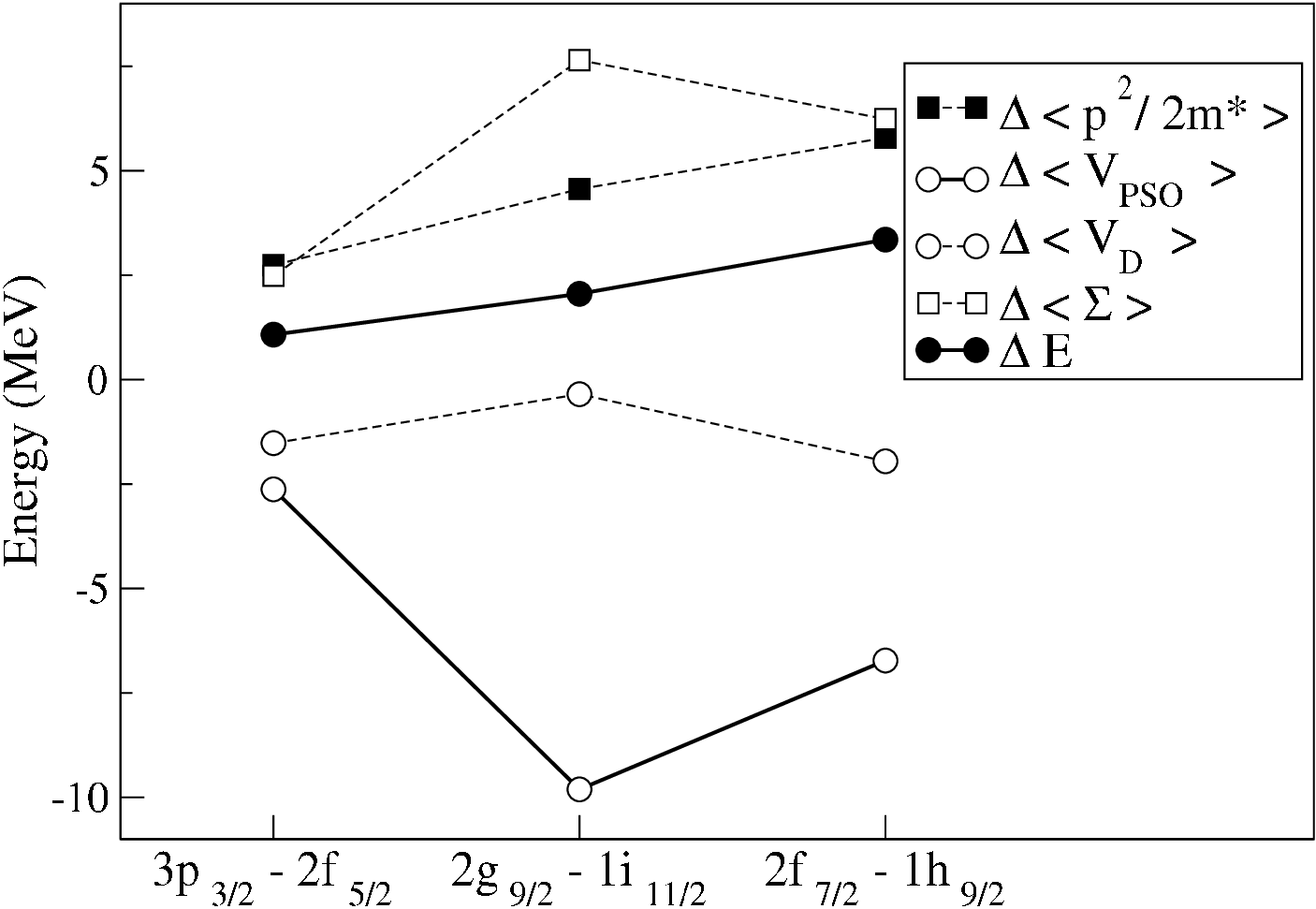}
\end{center}
\caption[Figure 1]{Differences for the energy terms in (\ref{aveg_schroed})
for $(1i_{11/2},\ 2g_{9/2})$, $(2f_{5/2},\ 3p_{3/2})$ and 
$(1h_{9/2},\ 2f_{7/2})$ pseudospin partners.}}
\hfill
\parbox[t]{5.5cm}{
\begin{center}
\includegraphics[clip=on,width=5.5cm]{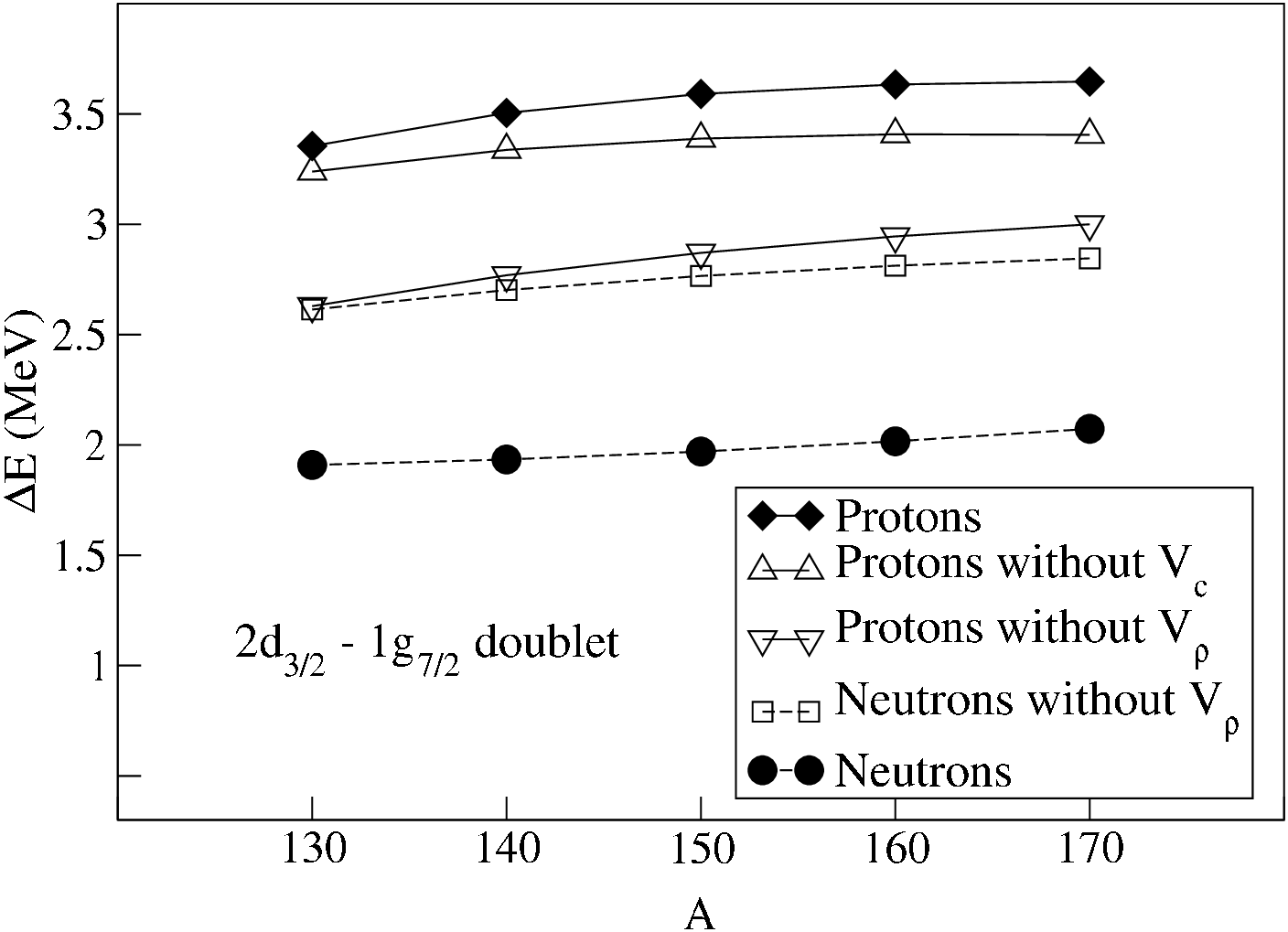}
\end{center}
\caption[Figure 2]{Effect of Coulomb and $\rho$ potentials $V_c$ and $V_\rho$ in
the splitting of the $(2d_{5/2},\ 1g_{7/2})$ pseudospin partners in a Sn isotopic
chain.}}
\end{figure}

In Fig.~1 are plotted the differences of 
the terms in Eq.~(\ref{aveg_schroed}) between 
each member of the three topmost pseudospin partners in
$^{208}$Pb, $(1i_{11/2},\ 2g_{9/2})$, 
$(2f_{5/2},\ 3p_{3/2})$ and $(1h_{9/2},\
 2f_{7/2})$. One sees that 
the contribution of $V_{\rm PSO}$ for the pseudospin energy splittings 
is larger than the splittings themselves and has the
opposite sign of the kinetic and $\langle\Sigma\rangle$ terms. For all
doublets there is a significant cancellation of 
$\big\langle V_{\rm PSO}\big\rangle$ with the kinetic and 
$\langle\Sigma\rangle$ terms. A similar analysis was made 
in ref.~\cite{marcos2}, concluding that the pseudospin-orbit 
potential is non-perturbative. We also found that there
is a clear correlation between $\big\langle V_{\rm PSO}\big\rangle$ 
and the pseudospin energy 
splitting when the diffusivity and the depth $\Sigma_0$ are varied
\cite{pmmdm_prl_prc}.

All these findings point to the conclusion that pseudospin symmetry has
a dynamical nature, since it is very much dependent on the shape of the 
$\Sigma$ nuclear mean field
potential and results from a strong cancellation 
among the several contributing terms to the energy splittings.

This pseudospin dependence on the shape of the $\Sigma$ potential allows us 
to explain why the proton
and neutron levels have different pseudospin splittings~\cite{meng}.

In RMF models, the $\rho$ meson interaction modifies the vector potential to
$
V = V_{\omega}+V_\rho=V_\omega\pm {g_{\rho}\over {2}} \rho_0
$,
where the plus sign refers to protons and the minus sign refers to neutrons, and
$\rho_0$ is the time component of the $\rho$ field which is proportional to $Z-N$.
The magnitude of the $\rho$ potential is around 4-8 MeV.
For heavy nuclei, $\rho_0<0$ and therefore $V_p<V_n$ which implies $|\Sigma_n| < |\Sigma_p|$.
This means that, for neutron-rich nuclei, the
pseudospin symmetry is favored for neutrons.
Notice, however, that the Coulomb potential, which has 
the opposite sign of $V_\rho$ for protons, also affects proton spectra. Both
of these effects can be seen in Figure 2, where is plotted the effect of
$V_\rho$ and the Coulomb potential on the energy splitting of the 
$(2d_{5/2},\ 1g_{7/2})$ 
pseudospin partners along the Sn isotopic chain from $A=130$ to $A=170$. We see that the
potential $V_\rho$ gives the main contribution for the observed isospin asymetry.
\end{section}

\vskip.4cm

\centerline{\bf Acknowledgments}
\vskip.3cm
We acknowledge financial support from FCT (POCTI), Portugal, and
from CNPq/ICCTI Brazilian-Portuguese scientific exchange program.


\begin{thebibliography}{99}
\bibitem{kth} K. T. Hecht and A. Adler, Nucl. Phys. {\bf A137}, 129 (1969)
\bibitem{aa} A. Arima, M. Harvey, and K. Shimizu, Phys. Lett. {\bf B30},
 517 (1969)
\bibitem{gino} J. N. Ginocchio, Phys. Rev. Lett. {\bf 78}, 436
(1997); {\textit ibid}, Phys. Rept. {\bf 315}, 231 (1999)
\bibitem{levi} J. N. Ginocchio and A. Leviatan, Phys. Lett.
{\bf B245}, 1 (1998)
 {\textit ibid}, Phys. Rev. Lett. {\bf 82}, 4599 (1999)
\bibitem{arima}  J. Meng, K. Sugawara-Tanabe, S. Yamaji,
 P. Ring, and A. Arima, Phys. Rev. {\bf C58}, R628 (1998)
\bibitem{marcos2} S. Marcos, M. L\'opez-Quelle, R. Niembro, L. N. Savushkin, and
P. Bernardos, Phys. Lett. {\bf B513}, 30 (2001)
\bibitem{pmmdm_prl_prc}P. Alberto, M. Fiolhais, M. Malheiro, A. Delfino, and
M. Chiapparini, Phys. Rev. Lett. {\bf 86}, 5015 (2001); {\textit ibid}, 
Phys. Rev. C65, 034307 (2002)
\bibitem{meng} J. Meng and I. Tanihata, Nucl. Phys. {\bf A650},
 176 (1999)
\end{thebibliography}
\end{document}